\documentclass[10pt, a4paper]{article}
\usepackage{lrec-coling2024}
\usepackage{graphicx}
\graphicspath{{./images/}}
\usepackage{tabularx}
\usepackage{float}
\usepackage{url}
\usepackage{hyperref}

\usepackage{inconsolata}
\usepackage{longtable}
\usepackage{enumitem}
\usepackage{multirow}
\usepackage{microtype}
\usepackage{amsfonts}
\newcolumntype{P}[1]{>{\centering\arraybackslash}m{#1}}
\usepackage{xcolor}
\definecolor{green}{RGB}{0, 128, 0}
\definecolor{red}{RGB}{196, 30, 58}

\usepackage[utf8]{inputenc}
\usepackage{algorithm}
\usepackage{algorithmic}
\usepackage{xfrac}
\usepackage{microtype}

\title{\textbf{ILCiteR:\\ Evidence-grounded Interpretable Local Citation Recommendation}}

\name{Sayar Ghosh Roy, Jiawei Han\\}

\address{University of Illinois Urbana-Champaign\\ \\ 
\texttt{\{sayar3, hanj\}@illinois.edu}\\}

\abstract{
Existing Machine Learning approaches for local citation recommendation directly map or translate a query, which is typically a claim or an entity mention, to citation-worthy research papers. Within such a formulation, it is challenging to pinpoint why one should cite a specific research paper for a particular query, leading to limited recommendation interpretability. To alleviate this, we introduce the evidence-grounded local citation recommendation task, where the target latent space comprises evidence spans for recommending specific papers. Using a distantly-supervised evidence retrieval and multi-step re-ranking framework, our proposed system, ILCiteR, recommends papers to cite for a query grounded on similar evidence spans extracted from the existing research literature. Unlike past formulations that simply output recommendations, ILCiteR retrieves ranked lists of evidence span and recommended paper pairs. Secondly, previously proposed neural models for citation recommendation require expensive training on massive labeled data, ideally after every significant update to the pool of candidate papers. In contrast, ILCiteR relies \textit{solely} on distant supervision from a dynamic evidence database and pre-trained Transformer-based Language Models without any model training. We contribute a novel dataset for the evidence-grounded local citation recommendation task and demonstrate the efficacy of our proposed conditional neural rank-ensembling approach for re-ranking evidence spans.
\\ \newline }

\begin{document}

\maketitleabstract

\section{Introduction}

\begin{figure}[t]
\includegraphics[width=0.482\textwidth]{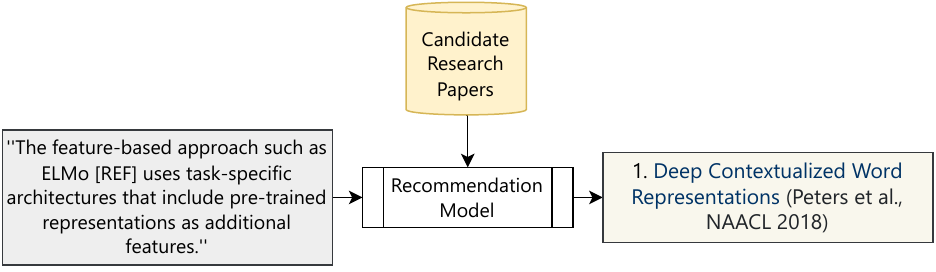}
\caption{An overview of the local citation recommendation task for scientific research papers.}
\label{fig:local_outline}
\end{figure}

A citation recommendation system retrieves a set of articles that could be cited for a given query. There are two broad types of citation recommendation tasks, namely, (a) global citation recommendation~\cite{who_to_cite}, and (b) local citation recommendation~\cite{neur_context_aware}. A global citation recommendation model fetches possible citations for a \textit{complete} document. In contrast, the local citation recommendation task aims to retrieve articles to cite conditioned on a much smaller text span (usually a sentence or a phrase). Fig.~\ref{fig:local_outline} provides a rough overview of the local citation recommendation task. Formally, the local citation recommendation task is defined as follows. Given a query $q$ and a set of $N$ candidate research papers $P = \{p_1, p_2, ..., p_N\}$, produce a ranked ordering: $[p_{i_1}, p_{i_2}, ..., p_{i_k}]$ of the top-$k$ ($k < N$) most relevant research papers that could be cited for the query $q$. The query usually represents a claim or an entity mention. For example, if $q$ = `ELMo', a valid citation could be \texttt{\{`title': `Deep Contextualized Word Representations', `year': 2018, ...\}}.

\begin{figure*}[t]
\begin{centering}
\includegraphics[width=1.00\textwidth]{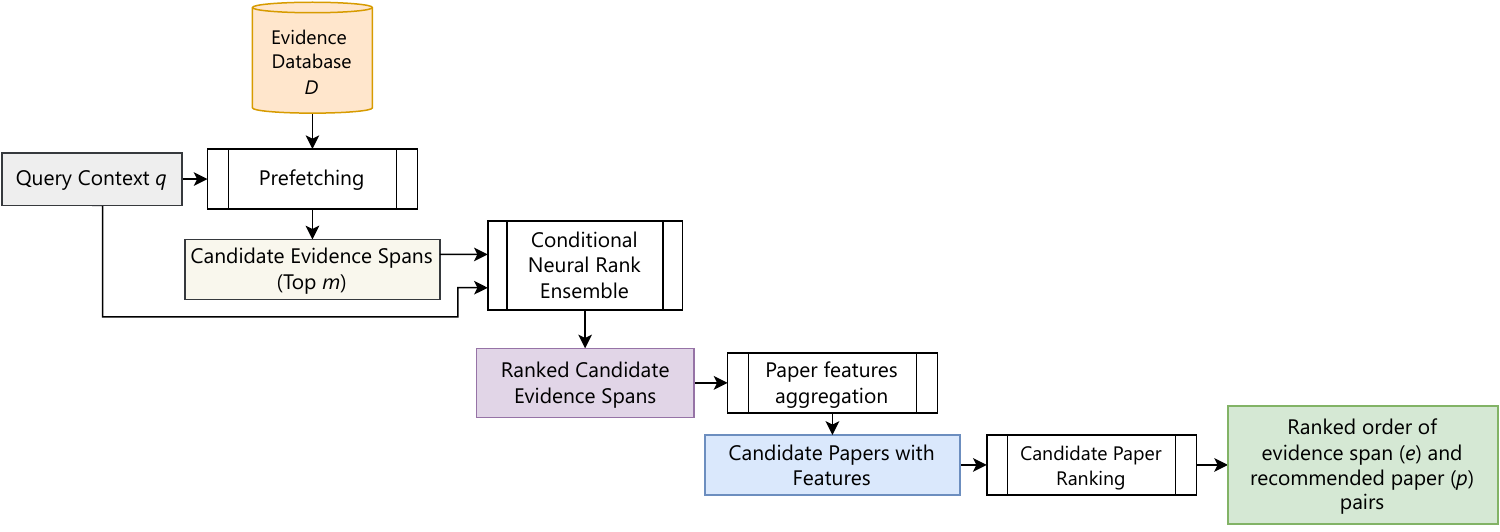}
\caption{An overview of ILCiteR: our proposed system for evidence-grounded citation recommendation.}
\label{fig:overview}

\end{centering}
\end{figure*}

In this work, we introduce the evidence-grounded local citation recommendation task shifting the focus to recommendation interpretability. Typical local citation recommendation methods estimate a \textit{direct} mapping function from the space of queries to that of research papers. Due to this formulation, there is no explicit reasoning mechanism to identify why a particular paper should be cited for a given query.

Consider the query $q$ = `NMT models trained on one domain tend to perform poorly when translating sentences in a significantly different domain'. Suppose, we find the following span $e$ in the existing literature: `it has also been noted that NMT models trained on corpora in a particular domain perform poorly when applied in a significantly different domain \texttt{[REF$_1$, REF$_2$]}' with \texttt{[REF$_1$]} referring to the paper $p_1$: \texttt{\{`title': `Six challenges for neural machine translation', `year': 2017, ...\}} and \texttt{[REF$_2$]} to $p_2$: \texttt{\{`title': `A survey of domain adaptation for neural machine translation', `year': 2018, ...\}}. We could then readily utilize $e$ as \textit{evidence} to recommend papers $p_1$ and $p_2$ to be cited for $q$. Intuitively, for a query $q$, a \textit{span} of text from an existing research paper which is both similar to $q$ and contains a citation for paper $p$, could readily serve as precedence to cite paper $p$ for query $q$.

Based on this principle, our proposed system, namely ILCiteR, retrieves \textit{evidence spans} ($e$'s) from the existing research literature that are (a) similar to the query, and (b) cite at least one paper. It re-ranks the retrieved evidence spans in decreasing order of their similarity to the query. It further ranks the individual papers associated with these ranked evidence spans. At the end of the 2-step re-ranking process, we obtain a ranked list of pairs of citation-worthy papers and appropriate evidence spans. Thus, every paper recommendation is grounded in the evidence span, which explicitly adds interpretability to each paper recommendation.

Our code and dataset are available here\footnote{\url{https://github.com/sayarghoshroy/ILCiteR}}. We make the following key contributions in this work.

\begin{enumerate}[leftmargin=16pt]
\setlength\itemsep{-1pt}
    
    \item We introduce the task of evidence-grounded local citation recommendation

    \item We present a novel dataset of over 200,000 unique evidence spans with corresponding sets of cited research paper and support pairs, covering three subtopics within Computer Science
    
    \item We propose ILCiteR, a distantly supervised approach for the evidence-grounded local citation recommendation task leveraging (I) an evidence database and (II) pre-trained Transformer language models, requiring \textit{no} explicit model training

    \item Our proposed conditional neural rank ensembling approach for re-ranking evidence spans significantly improves downstream paper recommendation performance over purely lexical and semantic similarity based retrieval as well as naive rank ensembling
    
\end{enumerate}

\section{Related Work}
\label{sec:rel_work}

\citet{context_aware_cr} defined the problem of local citation recommendation, laying out the distinction between global and local citation recommendation, and performed evaluations on CiteSeerX\footnote{\url{https://citeseerx.ist.psu.edu/}} using their prototype system. Inspired by the IBM Model 1~\cite{ibm_model} for Machine Translation, \citet{refseer} proposed a local citation recommendation system considering words within the query citation context as the source language and the list of candidate references as the target language.
Recently, various supervised Deep Learning models have been proposed for local citation recommendation~\cite{neur_context_aware,attention_local_cite,multi_info_fusion_cite,reco_het_bib_net,spr_smn}. \citet{improved_context_enhanced_global} enhanced the query by bringing in document level information (title and abstract of the paper from which the query originated). \citet{bert_gcn_citer} jointly considered the query's semantics and a citation network comprising various paper nodes. \citet{reco_scibert_hier} adopted a paper pre-fetching and re-ranking approach utilizing a Transformer-based Hierarchical-Attention text encoder and a SciBERT-based~\cite{scibert} paper ranking module. Other notable approaches include semantic modeling~\cite{semantic_cite} and recommendation ensembling~\cite{hybridcite}. \citet{semantic_cite} distinguish between two broad types of citations, namely citations for named entities and citations for claims. \citet{hybridcite} ensembles various existing methods such as LDA~\cite{LatentDA}, Doc2Vec~\cite{doc2vec}, Paper2Vec~\cite{paper2vec}, and HyperDoc2vec~\cite{hyperdoc2vec} using a semi-genetic recommender.

\vspace{6pt}

Our novel problem formulation for the \textit{evidence-grounded} recommendation task differs from all prior works on local citation recommendation. While past studies focused on building a neural mapping or translation model from the latent space of queries to that of paper nodes, our target latent space is that of evidence spans within the existing research literature.

As consequence of this updated formulation where the primary focus is interpretability, existing datasets~\cite{cite_reco_app_data,improved_context_enhanced_global,bert_gcn_citer,neur_context_aware} for local citation recommendation are unsuitable for our evidence-grounded recommendation task. We overcome this by building a comprehensive evidence database containing evidence spans and corresponding sets of cited papers with supports, and an evaluation set having queries extracted from recent papers never processed into the evidence database.

Lastly, our proposed system, ILCiteR, only leverages pre-trained Transformer-based language models with distant supervision from the dynamic evidence database and our neural recommendation pipeline does not require any form of training. Thus, we do not need to re-train ILCiteR after every update to the pool of candidate papers.

\section{Problem Definition}
\label{sec:prob_form_sec}

We formally define an evidence database and our task formulation as follows.

\paragraph{Evidence Database Specification}
\label{sec:ev_db_spec}

An evidence database ($D$) would contain individual records of the form: ($e, P$), where $e$ is an evidence span and $P$ is a set of $|P|$ pairs of the type $(p^j, s^j)$, $(j \in \{1, 2, ..., |P|\})$, with $p^j$ being the metadata for the $j^{th}$ paper in $P$ with $s^j$ representing its support, i.e., the number of times the paper $p^j$ was cited for the evidence span $e$. Each evidence span $e$ appears only once in the $D$.

The evidence database could also be regarded as map from evidence spans ($e$'s) to sets of paper metadata ($p$) and support ($s$) pairs. Table~\ref{tab:evidence_data_example} shows some examples of simplified records from an evidence database.

\paragraph{Task Formulation}
\label{sec:prob_form}

Given an evidence database $D$ and a query $q$, construct a ranked list of $n$ unique pairs of the form ($e^q_k$, $p^q_k$), $k \in \{1, 2, ..., n\}$, where $p^q_k$ is a recommended paper for $q$ with $e^q_k$ being the corresponding evidence span. The pairs are ranked in decreasing order of their relevance to $q$.\\

\section{Approach Overview}
\label{sec:approach_overview}

We briefly summarize our overall approach including evidence database creation and evidence-grounded paper recommendation.

To build the evidence database $D$ (Sec.~\ref{sec:ev_database}), we pre-process existing research papers to fetch individual sentences containing at least a single citation (a numbered \texttt{[REF]} tag). From each sentence, we extract the relevant portion of text, i.e., the evidence span, for each individual citation (Sec.~\ref{sec:evidence_context_extract}). Note that there might be multiple citations for a single evidence span. This comprehensive database $D$ acts as an external guide to provide evidence for citing a specific paper for some novel query.

For processing a query $q$, ILCiteR first pre-fetches a set of $m$ candidate evidence spans ($e$'s) from $D$ based on lexical similarity (Sec.~\ref{sec:prefetch}). It then utilizes a two-step re-ranking procedure to construct the ranked list of evidence span and recommended citation pairs (Sec.~\ref{sec:2-step-rank}). In the first step, it re-ranks the fetched evidence spans using a conditional neural rank ensembling approach (Sec.~\ref{sec:cond_rank_ens}). In the second step, it ranks the collection of all candidate papers corresponding to the $m$ retrieved evidence spans. The candidate papers are ranked considering the best observed rank of their associated evidence spans (from step one), their composite support, and recency (Sec.~\ref{sec:reco_papers}). An overview of our evidence-grounded local citation recommendation pipeline is shown in Fig.~\ref{fig:overview}.

\begin{table*}[t]
\begin{centering}
\begin{scriptsize}
\begin{tabular}{|m{5.8cm}|m{6.6cm}|P{1.28cm}|}

\hline
\textbf{Evidence span ($e$)}& \textbf{Primary cited paper metadata ($p$)} & \textbf{support ($s$)}\\
\hline
\hline
`fasttext'&\texttt{\{`title’: `Enriching word vectors with subword information’, `year’: 2016, `authors’: ...\}}&$34$\\
\hline
`Caffe'&\texttt{\{`title': `Caffe: Convolutional Architecture for Fast Feature Embedding', `year': 2014, `authors': ...\}}& $8$\\
\hline
`Gadag's method uses R-precision metric to evaluate the result of the paraphrased sentence'&\texttt{\{`title': `N-gram Based Paraphrase Generator from Large Text Document', `year': 2016, `authors': ...\}}&$1$\\
\hline
`demonstrated that AL outperformed random sampling for a simulated clinical NER task'&\texttt{\{`title': `A study of active learning methods for named entity recognition in clinical text', `year': 2015, `authors': ...\}}&$1$\\
\hline
`proposed to compress LSTM-based neural machine translation models with pruning algorithms'&\texttt{\{`title': `N-gram Based Paraphrase Generator from Large Text Document', `year': 2016, `authors': ...\}}&$1$\\
\hline

\end{tabular}
\end{scriptsize}

\caption{Examples of evidence span and primary cited paper with their corresponding support from the evidence database $D$. The first evidence spans contain simple entity mentions while the last three capture various types of claims.}
\label{tab:evidence_data_example}

\end{centering}
\end{table*}

\section{Evidence Database}
\label{sec:ev_database}

We build an evidence database $D$ to store evidence spans ($e$'s) and corresponding set of cited papers with support counts ($P$'s). Sec.~\ref{sec:ev_db_spec} lays out the specification for $D$. For our experiments, we consider three popular topics within Natural Language Processing, namely, Named Entity Recognition (\texttt{NER}), Summarization (\texttt{SUMM}), and Machine Translation (\texttt{MT}), and build evidence databases for each.

\subsection{Acquiring Research Papers}
\label{sec:acquire_papers}

We process the available dump of the \texttt{S2ORC} dataset\footnote{\url{https://github.com/allenai/s2orc}}~\cite{s2orc} to collect research papers tagged as `Computer Science'. To fetch papers on a particular topic, we check whether the topic name (or its abbreviation) appears within the paper abstract. Although simple, this method promises a high recall (as a work on the topic of say, Named Entity Recognition, is highly likely to have the topic name within the abstract itself). For building our evidence database, we only consider papers for which the full-text is \textit{publicly} available in some form and subsequently, its \textit{parsed} full-text exists within the \texttt{S2ORC} dump. We processed over 100 million papers from \texttt{S2ORC} and collected over 20,000 CS papers on the above topics (\texttt{2260} papers on \texttt{NER}, \texttt{8445} on \texttt{SUMM}, and \texttt{9522} on \texttt{MT}) having a publicly available full-text. From each collected paper $p$, we obtain a list $L$ of individual sentences containing \textit{at least} one citation. Thus, every sentence $s \in L$ has one or more \texttt{REF}s, each referring to a cited paper. We further process every $s$ to extract one or more evidence spans.

\begin{table}[]
\begin{centering}
\begin{scriptsize}
\begin{tabular}{|P{0.7cm}|P{0.8cm}|P{1.2cm}|P{1.3cm}|P{1.5cm}|}
\hline
topic&\#spans&avg \#chars&avg \#tokens&\#cited papers\\
\hline
\hline
\texttt{NER}&$23803$&$117.46$&$20.27$&$19041$\\
\hline
\texttt{SUMM}&$79345$&$118.89$&$20.37$&$59659$\\
\hline
\texttt{MT}&$108692$&$123.29$&$21.33$&$78743$\\       
\hline
\end{tabular}
\end{scriptsize}

\caption{Key statistics on the created evidence databases for $3$ popular CS subtopics}
\label{tab:data_stats}

\end{centering}
\end{table}

\begin{table*}[]
\begin{centering}
\begin{scriptsize}
\begin{tabular}{|m{2.4cm}|m{12.2cm}|}

\hline
Query $q_1$&`In the first step, we only look at one Transformer block, and describe how to learn the position representation driven by a dynamical system ...'\\
\hline
Retrieved Evidence $e_1$&`In the simplest case we use word embeddings and add position encodings to them; we use ...'\\
\hline
Recommendation $p_1$&\textcolor{green}{\texttt{\{`title': `Attention is all you need', `year': 2017, `authors': ...\}}}\\

\hline
\hline

Query $q_2$&`FastAlign'\\
\hline
Retrieved Evidence $e_1$&`FastAlign'\\
\hline
Retrieved Evidence $e_2$&`The alignment is induced with FastAlign'\\
\hline
Retrieved Evidence $e_3$&`One such method is the FastAlign word-alignment model'\\
\hline
Recommendation $p_1$&\textcolor{green}{\texttt{\{`title': `A simple, fast, and effective reparameterization of ibm model 2', `year': 2013, ...\}}}\\
\hline

\end{tabular}
\end{scriptsize}

\caption{Demonstration of our evidence-grounded paper recommendation system -- For a particular query, ILCiteR presents a ranked ordering of retrieved evidence span and recommended citation pairs.}
\label{tab:demo}

\end{centering}
\end{table*}

By using paper parses from \texttt{S2ORC}~\cite{s2orc}, the inherent noise from the \texttt{PDF} parsing process is minimized since, wherever applicable, \texttt{S2ORC} considers the available latex source leading to a more reliable parse. Also, the \texttt{[REF]} to cited paper mappings for our acquired papers are of high quality since \texttt{S2ORC} explicitly resolves bibliographic links among in-corpus paper clusters.

\subsection{Extracting Evidence Spans}
\label{sec:evidence_context_extract}

From a sentence $s$ containing at least one citation (a numbered \texttt{[REF]}), we extract \textit{relevant spans} for each \texttt{[REF]}, which serve as the evidence spans.
First, for every sentence $s \in L$, we group together sets of co-occuring \texttt{[REF]}s into unit \texttt{REFGROUP}s. For example, if $s$ = ``There are two broad types of text summarization approaches, namely, extractive [\texttt{REF}$_1$, \texttt{REF}$_2$, \texttt{REF}$_3$] and abstractive [\texttt{REF}$_4$]'', we would group \texttt{REF}s $1$ through $3$ into \texttt{REFGROUP}$_1$ and \texttt{REF}$_4$ into \texttt{REFGROUP}$_2$. Thus, \texttt{REFGROUP}$_1$ would refer to $3$ cited papers. The processed sentence $s'$ would be: ``There are two broad types of text summarization approaches, namely, extractive \texttt{REFGROUP}$_1$ and abstractive \texttt{REFGROUP}$_2$.'' We then extract evidence spans for each \texttt{REFGROUP} in $s'$.

We first isolate entity mentions in $s'$ by processing its dependency parse structure. We generate a dependency parse tree for $s'$ and begin a graph traversal from every \texttt{REFGROUP}-type node. Tokens visited during the traversal form the evidence span for the particular source \texttt{REFGROUP}. The traversal proceeds from a current node $curr$ to a child node $child$ only if:

\begin{enumerate}[leftmargin=16pt, itemsep=0pt]

    \item \texttt{pos}($curr$) $-$ \texttt{pos}($child$) = $1$, where \texttt{pos}($token$) is the index of $token$ in the processed sentence $s'$
    
    \item $\exists$ an edge in the produced dependency parse tree from $curr$ to $child$ encoding one the following relationships: `\texttt{COMP}' or `\texttt{AMOD}'

\end{enumerate}

Consider $s_1'$ = ``They used an IEX parser \texttt{REFGROUP}$_0$ to encode the ...'' and $s_2'$ = ``... past few years has been focused on extractive summarization \texttt{REFGROUP}$_2$ ...''. With the above rules, we capture `IEX Parser' as the evidence span with citation \texttt{REFGROUP}$_0$ and `extractive summarization' as that with citation \texttt{REFGROUP}$_2$.

In addition to dependency-parse based extraction, we extract evidence spans based on the sequence of tokens in $s'$. We split $s'$ into contiguous spans based on positions of \texttt{REFGROUP}s, and map each split span to its antecedent \texttt{REFGROUP}. Lastly, we use two additional conditions to include the entire sentence as a possible evidence span:

\begin{enumerate}[leftmargin=16pt, itemsep=0pt]

    \item The \texttt{REFGROUP} occurs at the sentence end. If $s_3'$ = ``Context embeddings were generated using Sentence Transformers \texttt{REFGROUP}$_0$'', \texttt{REFGROUP}$_0$ could either be a citation for (a) `Sentence Transformers' or (b) the complete sentence
    
    \item Only \textit{one} \texttt{REFGROUP} exists within $s'$. If $s_4'$ = ``They used ROUGE and METEOR metrics \texttt{REFGROUP}$_0$ for evaluating their models'', it would be beneficial to consider the full sentence as an evidence span for \texttt{REFGROUP}$_0$. 
    
\end{enumerate}

Note that we apply the above dependency parse and token sequence based extraction rules in a hierarchical fashion. As an example, for $s_5'$ = ``They used BERT \texttt{REFGROUP}$_0$, a popular Large Language Model \texttt{REFGROUP}$_1$, to generate text embeddings \texttt{REFGROUP}$_2$'', we have the following evidence spans ($e$'s) and their corresponding sets of cited papers (captured within the \texttt{REFGROUP}s):

\begin{itemize}[leftmargin=12pt]
    \setlength\itemsep{-4pt}
    \item[] `BERT' $\rightarrow$ \texttt{REFGROUP}$_0$
    \item[] `Large Language Model' $\rightarrow$ \texttt{REFGROUP}$_1$
    \item[] ``to generate text embeddings'' $\rightarrow$ \texttt{REFGROUP}$_2$
    \item[] Complete Sentence $\rightarrow$ \texttt{REFGROUP}$_2$
\end{itemize}

\subsection{Populating Evidence Database}
\label{sec:build_D}

We process every sentence (containing at least one citation) from each collected paper to extract evidence spans ($e_i$'s), $i \in \{1, 2, ..., |D|\}$ and corresponding \texttt{REFGROUP}s. We then resolve \texttt{REFGROUP}s into their component \texttt{REF}s. We populate the evidence database $D$ using pairs of $e_i$'s and sets of pairs of cited paper ($p_i^j$) and support ($s_i^j$), $j \in \{1, 2, ..., |P_i|\}$. We fetch papers within $P_i$ from the resolved \texttt{REF}s' metadata in \texttt{S2ORC}. The support count $s_i^j$ contains the number of times we observed $p_i^j$ being cited for the evidence span $e_i$. Within $D$, $e_i$'s are unique. Therefore, $D$ could be regarded as a map from evidence spans ($e$'s) to sets to cited paper and support pairs ($P$'s). During construction, we do not perform any additional semantic conflation of $e_i$'s; only occurrences of matching evidence spans are continually conflated.

\subsection{Dataset Statistics}
\label{sec:db_stats}

We provide key statistics of our evidence databases created for the $3$ CS subtopics in table~\ref{tab:data_stats}. Also, as an illustration, we present some evidence spans and their primary cited article with corresponding support in table~\ref{tab:evidence_data_example}. Here, we only show the most prominent cited article from the set $P$ with its support. The first two evidence spans in table~\ref{tab:evidence_data_example} are simple entity mentions while the last three represent various types of claims. We make our evidence databases (and evaluation splits) publicly available\footnote{\url{https://github.com/sayarghoshroy/ILCiteR}}.

\begin{figure}[]
\begin{centering}
\includegraphics[width=0.476\textwidth]{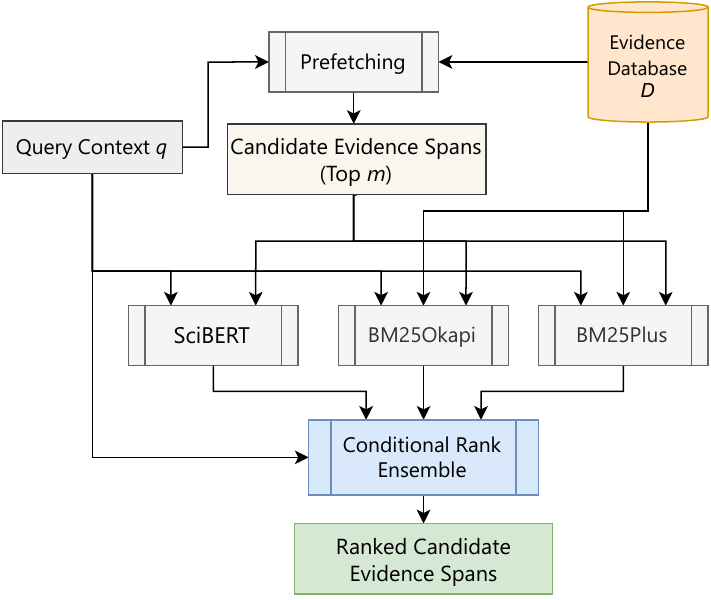}
\caption{Conditional neural rank ensembling -- re-rank candidate evidence text spans based on lexical and semantic similarity to the query $q$.}
\label{fig:cand_rerank}
\end{centering}
\end{figure}

\section{Evidence-grounded Recommendation}
\label{sec:method}

In this section, we describe our overall methodology for evidence-grounded paper recommendation. Our task formulation can be found in Sec.~\ref{sec:prob_form_sec}. Given a query $q$, we first pre-fetch a set of $m$ candidate evidence spans from $D$ (Sec.~\ref{sec:prefetch}). We then adopt a conditional neural rank ensembling approach to re-rank pre-fetched evidence spans based on their similarity to $q$ (Sec.~\ref{sec:cond_rank_ens}). In that, we conditionally ensemble a ranking of evidence spans based on semantic similarity only for lengthier queries.
Lastly, we rank the collection of candidate papers based on both the relevance of their underlying evidence span and other paper-specific features (Sec.~\ref{sec:reco_papers}). An overview of our recommendation methodology, namely ILCiteR, is presented in Figure.~\ref{fig:overview}.

% 0.00420, 0.00363, 0.00385
% 0.004, 0.006, 0.004

\begin{table*}[]
\begin{centering}
\begin{scriptsize}
\begin{tabular}{|c|c|c|c|c|c|c|}
\hline
Topic                & Method                & MRR              & R@1            & R@3            & R@5            & R@10           \\ \hline
\multirow{3}{*}{\texttt{NER}}  & BM25Okapi              & $0.34558$          & $0.260$          & $0.380$          & $0.430$          & $0.512$          \\ 
                      & BM25Plus               & $0.34735$          & $0.262$          & $0.386$          & $0.436$          & $0.510$          \\ 
                      & Conditional Ensembling & $\textbf{0.35155}$ & $\textbf{0.266}$ & $\textbf{0.390}$ & $\textbf{0.438}$ & $\textbf{0.514}$ \\ \hline
\multirow{3}{*}{\texttt{SUMM}} & BM25Okapi              & $0.37052$          & $0.282$          & $0.410$          & $0.458$          & $0.530$          \\ 
                      & BM25Plus               & $0.37324$          & $0.284$          & $0.410$          & $\textbf{0.460}$ & $0.540$          \\ 
                      & Conditional Ensembling & $\textbf{0.37687}$ & $\textbf{0.290}$ & $\textbf{0.414}$ & $0.458$          & $\textbf{0.542}$ \\ \hline
\multirow{3}{*}{\texttt{MT}}   & BM25Okapi              & $0.43568$          & $0.332$          & $0.506$          & $0.546$          & $0.620$          \\ 
                      & BM25Plus               & $0.44071$          & $0.338$          & $0.500$          & $\textbf{0.558}$ & $\textbf{0.636}$ \\
                      & Conditional Ensembling & $\textbf{0.44456}$ & $\textbf{0.342}$ & $\textbf{0.516}$ & $0.556$          & $0.628$          \\ \hline
\end{tabular}
\end{scriptsize}

\caption{Performance of proposed conditional neural rank ensembling approach and lexical similarity based baselines. In general, conditional ensembling \textit{significantly} improves recommendation performance.}
\label{tab:main_results}

\end{centering}
\end{table*}

\begin{table*}[]
\begin{centering}
\begin{scriptsize}
\begin{tabular}{|c|c|c|c|c|c|c|}
\hline
Topic                & Method                      & MRR              & R@1            & R@3            & R@5            & R@10           \\ \hline
\multirow{5}{*}{\texttt{NER}}  & SciBERT only                 & $0.26246$          & $0.188$          & $0.268$          & $0.342$          & $0.428$          \\ 
                      & Ensemble BM25Okapi + SciBERT & $0.31272$          & $0.228$          & $0.334$          & $0.404$          & $0.506$          \\ 
                      & Ensemble BM25Plus + SciBERT  & $0.30772$          & $0.220$          & $0.338$          & $0.400$          & $0.496$          \\ 
                      & Ensemble BM25s + SciBERT     & $0.31502$          & $0.222$          & $0.344$          & $0.418$          & $0.510$          \\ 
                      & Conditional Ensemble         & $\textbf{0.35155}$ & $\textbf{0.266}$ & $\textbf{0.390}$ & $\textbf{0.438}$ & $\textbf{0.514}$ \\ \hline
\multirow{5}{*}{\texttt{SUMM}} & SciBERT only                 & $0.31555$          & $0.234$          & $0.346$          & $0.392$          & $0.496$          \\ 
                      & Ensemble BM25Okapi + SciBERT & $0.34795$          & $0.266$          & $0.374$          & $0.414$          & $0.520$          \\ 
                      & Ensemble BM25Plus + SciBERT  & $0.35251$          & $0.274$          & $0.370$          & $0.408$          & $0.526$          \\ 
                      & Ensemble BM25s + SciBERT     & $0.35822$          & $0.270$          & $0.392$          & $0.438$          & $0.530$          \\ 
                      & Conditional Ensemble         & $\textbf{0.37687}$ & $\textbf{0.290}$ & $\textbf{0.414}$ & $\textbf{0.458}$ & $\textbf{0.542}$ \\ \hline
\multirow{5}{*}{\texttt{MT}}   & SciBERT only                 & $0.38589$          & $0.288$          & $0.430$           & $0.488$          & $0.560$          \\ 
                      & Ensemble BM25Okapi + SciBERT & $0.42961$          & $0.342$          & $0.458$          & $0.514$          & $0.614$          \\ 
                      & Ensemble BM25Plus + SciBERT  & $0.44116$          & $0.352$          & $0.472$          & $0.542$          & $0.630$          \\ 
                      & Ensemble BM25s + SciBERT     & $0.44305$          & $\textbf{0.356}$ & $0.474$          & $0.540$          & $\textbf{0.636}$ \\ 
                      & Conditional Ensemble         & $\textbf{0.44456}$ & $0.342$          & $\textbf{0.516}$ & $\textbf{0.556}$ & $0.628$          \\ \hline
\end{tabular}
\end{scriptsize}

\caption{Metrics achieved by re-ranking using SciBERT-based semantic similarity only, naive ensemble of lexical and semantic similarity ranks, and our proposed conditional neural rank ensembling approach.}
\label{tab:ablations}

\end{centering}
\end{table*}

\subsection{Pre-fetching}
\label{sec:prefetch}

To reduce the search space of evidence spans, we first prefetch a set of $\le 100$ candidate evidence spans from the evidence database $D$ that have the highest lexical similarity to the query $q$. These candidate evidence spans are then re-ranked in order of their similarity to the query $q$ during the conditional rank ensembling step (Sec.~\ref{sec:cond_rank_ens}).

Given a query text span $q$, we score each evidence span ($e$) in the evidence database $D$ using Okapi-BM25~\cite{bm25} and BM25Plus~\cite{bm25plus}. We formally write out the Okapi-BM25 score and the BM25plus score as follows.

\paragraph{Okapi-BM25} Given a query $q$ with tokens $q_1$, $q_2$, ..., $q_l$, the Okapi-BM25 score for a particular evidence span $e$ can be written as: \texttt{BM25Okapi\_score}$(e, q) = \sum_{i = 1}^{l} $\texttt{IDF}$(q_i) \cdot \frac{f(q_i, e) \cdot (k_1 + 1)}{f(q_i, e) + k_1 \cdot (1 - b + b \cdot \frac{|e|}{a})}$. Here, $f(q_i, e)$ equals the number of times the token $q_i$ is found within the evidence span $e$, $|e|$ equals the number of tokens within the evidence span $e$, and $a$ is the average token length of evidence spans within the evidence database $D$. $k_1$ and $b$ are hyperparameters.

Also, \texttt{IDF}$(q_i) = ln(\frac{|D| - n(q_i) + 0.5}{n(q_i) + 0.5} + 1)$, where $n(q_i)$ represents the number of evidence spans within $D$ that contain $q_i$.

\paragraph{BM25plus} BM25Plus has an additional free parameter, $\delta$, with a default value of $1.0$. The BM25Plus score can be written as: \texttt{BM25Plus\_score}$(e, q) = \sum_{i = 1}^{l} $\texttt{IDF}$(q_i) \cdot$ $\left( \frac{f(q_i, e) \cdot (k_1 + 1)}{f(q_i, e) + k_1 \cdot (1 - b + b \cdot \frac{|e|}{a})} + \delta \right)$.

With standard BM25Okapi~\cite{bm25}, the term frequency normalization by the length of the evidence span would not be perfectly lower bounded. Therefore, Okapi BM25 would tend to score longer evidence spans unfairly as having a relevancy score similar to that of shorter evidence spans not containing the query tokens. The BM25Plus algorithm~\cite{bm25plus} resolves this drawback by introducing the parameter $\delta$.

We consider an evidence span from $D$ as a \textit{candidate} evidence span for query $q$ if it was among the top-$50$ highest scored evidence spans by either BM25Okapi or BM25Plus. We found that pre-fetching $50$ evidence spans for each of BM25Okapi and BM25Plus provided a reasonably high recall and the effects of increasing this threshold beyond $50$ were negligible. We thus create a set of $m$ candidate evidence spans $E^q = \{e^q_1, e^q_2, ..., e^q_m\}$ from $D$.

Note that the pre-fetching step in our evidence-grounded citation recommendation task differs from traditional pre-fetching in past works on local citation recommendation~\cite{reco_scibert_hier,neur_context_aware}. Earlier works directly pre-fetched candidate \textit{research articles}. In contrast, we fetch \textit{evidence spans} that are relevant to the query $q$.

\subsection{Two-step Ranking}
\label{sec:2-step-rank}

We first re-rank evidence spans obtained from the pre-fetching step using our conditional neural rank ensembling approach and subsequently rank the collection of candidate papers.

\subsubsection{Conditional Rank Ensembling}
\label{sec:cond_rank_ens}

We formulate a conditional neural rank ensembling method to re-rank pre-fetched candidate evidence spans (evidence spans in $E^q = \{e^q_1, e^q_2, ..., e^q_m\}$). Our design choice is based on the following observation. For smaller and simple queries, utilizing overlap of key lexical items is important since shorter queries typically contain mentions of notable named entities. However, when the query is long and complex, effectively capturing its semantics becomes crucial. To this end, for short queries, we use a rank ensemble of BM25Plus~\cite{bm25plus} and BM25Okapi~\cite{bm25}. When the query length exceeds a threshold, we leverage a pre-trained SciBERT~\cite{scibert} model. SciBERT is a Transformer-based~\cite{transformer} Large Language Model (similar to BERT~\cite{bert}) pre-trained on large corpora of text from scientific research papers. We define the semantic similarity score for $e^q_i$ ($i \in \{1, 2, ..., m\}$) as the cosine similarity between the $768$-dimensional \texttt{[CLS]} embeddings of the query $q$ and $e^q_i$.

Formally, if $E_q$ and $E_e$ are the SciBERT generated embeddings of the initial \texttt{[CLS]} tokens of the input query $q$ and a candidate evidence span $e$, the semantic similarity score between $q$ and $e$ could be written as: \texttt{semantic\_score}$(q, e) = \frac{E_q \cdot E_e}{|E_q| \cdot |E_e|}$, where $E_q$, $E_e$ $\in \mathbb{R}^{768}$.

We ensemble the SciBERT-based semantic similarity rank with the BM25Plus rank to produce the ranked ordering for lengthier queries. We thus obtain an evidence rank $r^q_i$ for each $e^q_i$. At the end of this step, we sort $e^q_i$'s based on $r^q_i$'s in ascending order and pass the rank-sorted order: [$e^q_i$, $e^q_i$, ..., $e^q_i$] to the final paper recommendation step. A rough overview of our conditional rank ensembling approach is shown in Fig.~\ref{fig:cand_rerank}.

% SciBERT is a Transformer-based~\cite{transformer} Large Language Model (similar to BERT~\cite{bert}) pre-trained on large corpora of text from scientific research papers.

\begin{table*}[]
\begin{centering}
\begin{scriptsize}
\begin{tabular}{|m{2.4cm}|m{12.4cm}|}

\hline
Query $q$&`exploit monolingual data in two ways: through self-learning by forward-translating the monolingual source data to create synthetic parallel data ...'\\
\hline
\hline
Method&BM25 ensemble (BM25Okapi + BM25Plus)\\
\hline
Retrieved Evidence $e_1$&`By applying a self-training scheme, the pseudo parallel data were obtained by automatically translating the source-side monolingual corpora ...'\\
\hline
Model Reco. $p_1$&\textcolor{red}{\texttt{\{`title': `Transductive learning for statistical machine translation', `year': 2007, ...\}}}\\

\hline
\hline

Method&Conditional Neural Rank Ensembling\\
\hline
Retrieved Evidence $e_1$&`... employing a self-learning algorithm to generate pseudo data, while the second is using two NMT models to predict the translation and to reorder the source-side monolingual sentences'\\
\hline
Model Reco. $p_1$&\textcolor{green}{\texttt{\{`title': `Exploiting source-side monolingual data in neural machine translation', `year': 2016, ...\}}}\\
\hline

\end{tabular}
\end{scriptsize}

\caption{Conditionally utilizing semantic similarity scoring (using SciBERT) for lengthier contexts helps retrieve accurate evidence text spans leading to better recommedations.}
\label{tab:condi_helps}

\end{centering}
\end{table*}

\begin{table*}[t]
\begin{scriptsize}
\begin{tabular}{|m{2.4cm}|m{12.2cm}|}

\hline
Query $q_1$&`The synchronous context-free grammars introduced by are transduction grammars whose productions have some restrictions'\\
\hline
Retrieved Evidence $e_1$&`The model is based on inversion transduction grammars (ITGs), a variety of synchronous context free grammars (SCFGs)'\\
\hline
Retrieved Evidence $e_2$&`Among the grammar formalisms successfully put into use in syntax based SMT are synchronous context-free grammars (SCFG) and synchronous tree substitution grammars (STSG)'\\
\hline
Model Reco. $p_1$&\textcolor{red}{\texttt{\{`title': `Stochastic inversion transduction grammars and bilingual parsing of parallel corpora', `year': 1997, ...\}}}\\
\hline
True Reco. $p_g$&\textcolor{green}{\texttt{\{`title': `Hierarchical phrase-based translation', `year': 2007, ...\}}}\\
\hline

\end{tabular}
\end{scriptsize}

\caption{Instance of an error made by our model. Though the ground truth recommendation $p_g$ is more appropriate, evidence spans retrieved by our model are relevant, $p_1$ is a sensible recommendation for $q_1$.}
\label{tab:error_1}
\end{table*}

\begin{table*}[]
\begin{scriptsize}
\begin{tabular}{|m{2.4cm}|m{12.2cm}|}

\hline
Query $q_2$&`bidirectional long short-term memory with conditional random field model (BiLSTM-CRF) exhibited promising results'\\
\hline
Retrieved Evidence $e_1$&`Among others, the model of bidirectional Long Short Term Memory with a conditional random field layer (BiLSTM-CRF), exhibits promising results'\\
\hline
Retrieved Evidence $e_2$&`which is based on a bidirectional long short-term memory (LSTM) network with a conditional random field (CRF) over the output layer'\\
\hline
Retrieved Evidence $e_3$&`Our framework is based on a bidirectional long short-term memory network with a conditional random fields (BiLSTM-CRFs) for sequence tagging'\\
\hline
Model Reco. $p_1$&\textcolor{red}{\texttt{\{`title': `Neural architectures for named entity recognition', `year': 2016, `venue': `NAACL', ...\}}}\\
\hline
Model Reco. $p_2$&\textcolor{red}{\texttt{\{`title': `Bidirectional LSTM-CRF models for sequence tagging', `year': 2015, ...\}}}\\
\hline
True Reco. $p_g$&\textcolor{green}{\texttt{\{`title': `End-to-end sequence labeling via bi-directional lstm-cnns-crf', `year': 2016, `venue': `ACL', `authors': ...\}}}\\
\hline

\end{tabular}
\end{scriptsize}

\caption{Instance of an error made by our model -- though the ground truth recommendation $p_g$ does not match the primary model recommendation $p_1$, $p_1$ is an appropriate recommendation for $q_1$ as it was cited for claims that are very similar to $q_2$.}
\label{tab:error_2}
\end{table*}

\begin{table*}[]
\begin{centering}
\begin{scriptsize}
\begin{tabular}{|m{2.25cm}|m{12.35cm}|}

\hline
Query $q_3$&`The use of tokens to condition the output of NMT started with the multilingual models'\\
\hline
Retrieved Evidence $e$&`... the idea to multilingual NMT by concatenating parallel data of various language pairs and marking the source ...'\\
\hline
Model Reco. $p$&\textcolor{red}{\texttt{\{`title': `Google's multilingual neural machine translation system: Enabling zero-shot translation', `year': 2017, `venue': `TACL', ...\}}}\\
\hline
True Reco. $p_g$&\textcolor{green}{\texttt{\{`title': `Google's multilingual neural machine translation system: Enabling zero-shot translation', `year': 2016, ...\}}}\\
\hline

\end{tabular}
\end{scriptsize}

\caption{Recommendation error due to multiple versions of the same paper.}
\label{tab:error_3}

\end{centering}
\end{table*}

\subsubsection{Recommending Papers with Evidence}
\label{sec:reco_papers}

From the previous evidence re-ranking step, we obtain the ranked order of $m$ candidate evidence text spans for query $q$: $E^q = [e^q_1, e^q_2, ..., e^q_m]$, with $P^q_i$ being the set of pairs of research paper and support associated to $e^q_i$ obtained from the evidence database $D$ ($i \in \{1, 2, ..., m\}$). Here, for any $j < k$, $e^q_j$ would be more similar to $q$ than $e^q_k$.

For each $e^q_i$, we obtain the associated set of $|P^q_i|$ cited papers with their supports from $D$:

\begin{center} $P^q_i = \{(_1p^q_i,$ $_1s^q_i), (_2p^q_i,$ $_2s^q_i), ..., (_{|P^q_i|}p^q_i,$ $_{|P^q_i|}s^q_i)\}$ \end{center}

With $_js^q_i$ being the support for paper $_jp^q_i$ in $P^q_i$, $j \in \{1, 2, ..., |P^q_i|\}$. We now rank the \textit{set} of all papers among pairs in all $P^q_i$'s, accounting for: (1) the highest observed rank of a particular paper's associated evidence span, (2) the paper's composite support, and (3) its recency. We formally define each of these components as follows:

\begin{enumerate}[leftmargin=16pt]
\setlength\itemsep{0pt}

\item The lowest observed rank of a paper $p$:

\begin{centering}

$p_r = min_{i \in \{1, 2, ..., m\}}\ R_i(p)$,

\end{centering}

where $R_i(p) = i$ if $\exists$ $s$ $s.t.$ $(p, s) \in P^q_i$, else $R_i(p) = \infty$

\item The total support for a paper: 

\begin{centering}

$p_s = \sum_{i = 1}^{m} S_i(p)$

\end{centering}

where $S_i(p) = s$ if $\exists$ $s$ $s.t.$ $(p, s) \in P^q_i$, else $S_i(p) = 0$

\item The recency of a paper ($rec_p$), which in our case, simply holds the year of publication 

\end{enumerate}

We rank papers in the following order of precedence: (1) $p_r$ (lower is better), (2) $p_s$ (higher is better), (3) $rec_p$ (higher is better).

For every evidence span, we obtain a ranked order of cited papers. We aggregate this into a ranked ordering of evidence span and recommended citation pairs.
Thus, each paper recommendation for a query $q$ is grounded in the accompanying evidence span, which provides explicit interpretability.
Our evidence database could be dynamically up-scaled to include evidence spans from newly published papers.

\section{Experiments and Analysis}
\label{sec:expt_analysis}

We evaluate the performance of our evidence-grounded recommendation system and present some key insights.

\subsection{Evaluation Set Creation}
\label{sec:eval_set_create}

For a particular topic, we first create an evaluation candidates set $E_{c}$ containing candidate queries and their ground truth cited papers extracted\footnote{Applying the procedure detailed in Sec.~\ref{sec:evidence_context_extract}} from the $200$ most recent papers on that topic which were not considered while creating $D$. This ensures that $D$ does not contain any mappings extracted from papers that contribute to the evaluation set, reproducing a real-world evaluation setting. We then create our final evaluation set $E$ by picking $500$ datapoints ($d$'s) from $E_{c}$ satisfying the following condition:

\vspace{4pt}

For some $d = (q, \{p'_1, p'_2, ..., p'_a\}) \in E$, there is at least one pair of evidence span $e$ and a set $P$ of pairs of cited paper and support, with $P = \{(p^1, s^1), (p^2, s^2), ..., (p^b, s^b)\}$) such that $(e, P) \in D$ and for some $(i, j)$ with $i \in \{1, 2, ..., a\}$ and $j \in \{1, 2, ..., b\}$, $p'_i = p^j$. Here, $q$ is the query and $\{p'_1, p'_2, ..., p'_a\}$ is the set of ground truth cited papers for $q$.

\vspace{4pt}

This ensures that for every datapoint $d$ in $E$, at least one ground truth paper exists within the set of all candidate papers in $D$.

\subsection{Interpretable Paper Recommendation}
\label{sec:paper_reco}

We show two examples to demonstrate the working of our evidence-grounded local citation recommendation system (ILCiteR) in table~\ref{tab:demo}. In table~\ref{tab:main_results}, we compare the performance of our proposed conditional neural rank ensembling approach for re-ranking evidence spans with lexical similarity based baselines. We use the Mean Reciprocal Rank (MRR) and Recall@$N$ measures (with $N = 1, 3, 5, 10$) as automatic evaluation metrics.

From table~\ref{tab:demo}, for $q_1$, we see that our system extracts a highly relevant evidence span (describing the Transformer model with a focus on position embeddings). $q_2$ demonstrates that handling entity mentions is relatively straightforward for our system -- several pieces of evidence spans from existing papers are provided to suggest why $p_2$ should be cited for the entity mention `FastAlign'.

From table~\ref{tab:main_results}, we see that our conditional neural rank ensembling approach improves recommendation performance over BM25Okapi and BM25Plus on Recall@$1$, Recall@$3$, and MRR, for all topics. Overall, we see performance improvements on all metrics with conditional rank ensembling with the exception of R@$5$ for \texttt{SUMM}, \texttt{MT} and R@$10$ for \texttt{MT} (where lexical similarity using BM25Plus leads to better recommendations). We performed t-tests which showed that our proposed conditional ensembling approach improves scores significantly over unsupervised BM25Okapi's ranking with a $p$-value of $0.08$\footnote{An exception being Recall@$5$ for \texttt{topic: NER} where conditional ensembling led to slightly worse results.} for MRR and Recall@$5$ metrics.

\subsection{Ablation Studies}
\label{sec:ablations}

We perform several ablations to showcase the efficacy of our proposed conditional neural rank ensembling approach. In table~\ref{tab:ablations}, we present metrics for re-ranking using SciBERT-based semantic similarity only, naive ensemble of lexical and semantic similarity ranks, and our proposed conditional rank ensembling. Firstly, we notice that re-ranking with plain SciBERT massively drops recommendation performance. This suggests that lexical similarity is a useful signal for evidence span re-ranking. Secondly, we observe that our conditional ensembling approach consistently produces performance improvements on all metrics across all topics (R@$1$ and R@$5$ for \texttt{MT} are the only exceptions where naive rank ensembling using SciBERT and lexical similarity achieves better numbers). We performed t-tests which showed that our proposed conditional rank ensembling approach significantly improved model performance over re-ranking evidence spans purely using SciBERT, and a naive ensemble of SciBERT and BM25, with a $p$-value of $5\%$\footnote{An exception being \texttt{topic: MT} where the naive ensemble's MRR was similar to our proposed approach.}. In general, unconditionally ensembling with semantic similarity for queries of all lengths worsens performance compared to using semantic similarity ranks only when necessary, i.e., when the query is lengthy enough. We observed that a length threshold $\approx 2.5$ times the average token length for evidence spans ($\approx 50$ tokens) was optimal for ensembling with semantic similarity-based ranks.

In table~\ref{tab:condi_helps}, we present a concrete example of how conditionally leveraging semantic similarity scoring (using SciBERT) for lengthy contexts helps evidence span retrieval. For $q$, we are able to extract a moderately relevant evidence span using ensembled lexical similarity. But although the suggested evidence span shares multiple common entities with the query, the downstream recommendation is incorrect (recommends a paper on statistical MT). Without the semantic similarity based signal, the ground truth paper does not appear among the top-$3$ recommendations. Here, conditional rank ensembling captures the semantics of $q$ and accurately retrieves an evidence span pointing to the ground truth paper on neural machine translation.

\subsection{Error Analysis}
\label{sec:error}

Consider the query $q_1$ from table~\ref{tab:error_1}. The top-$2$ evidence spans retrieved by our model were $e_1$ and $e_2$ with the recommended citation being $p_1$, which was incorrect. We observe that the retrieved evidences are relevant to $q_1$, and $p_1$ could possibly be cited for such a context. However, the ground truth citation $p_G$ is more appropriate for $q_1$, especially since the underlying topic is \texttt{MT}. Deciding whether to cite a paper on the method itself (in this case, Stochastic ITGs) or a paper on how the method is applied to a specific task (here, Machine Translation) would require additional input fields like the exact topic or area of the research paper from which the query originates. Handling such cases is tricky for our system since we have query as the \textit{only} input field.
Note that our model suggests $p_g$ as recommendation \#$9$ with the following evidence span: `use some grammar at their core, for instance: (a) synchronous context-free grammars (SCFG)'.

We have a similar case in table~\ref{tab:error_2}. Here, our model retrieves highly relevant evidence spans ($e_1$, $e_2$, and $e_3$) and recommends papers $p_1$ and $p_2$. We observe that the evidence span $e_1$ is very similar to $q_2$ and $p_1$ is indeed a valid citation for $q_2$. Here, the ground truth citation points to a paper on the more general use of LSTM-CNNs-CRFs (in end-to-end sequence-labeling) while $p_1$ is a paper on the topic of interest, i.e., neural models for Named Entity Recognition, from the same year. $p_2$, which is ranked second by the model, is very similar to $p_g$. And one could even argue that since $p_2$ is about LSTM-CRFs for sequence tagging, it is more appropriate for $q_2$ compared to $p_g$ which is a work on using LSTM-CNNs-CRFs for general sequence labeling. Thus, we observe that even when the model recommendations for a query do not match the ground truth, the retrieved evidence spans are highly relevant and the model recommended papers are quite appropriate.

We also find another cause of error due to multiple versions of the same paper (Table~\ref{tab:error_3}). The model recommendation $p$ was deemed as a failure (as we do not conflate paper nodes from different years) since the ground truth citation was a pre-print version of the same paper from an earlier year.

\subsection{Implementation Details}
\label{sec:implement_details}

We use publicly available implementations of BM25\footnote{\url{https://github.com/dorianbrown/rank_bm25}}, with default values for all hyperparameters. For generating dependency parses, we utilized \texttt{spaCy}'s\footnote{\url{https://spacy.io/}} default dependency parser based on \citet{spacy_parser} and \citet{dep_pseudo}. We used \texttt{HuggingFace}'s\footnote{\url{https://huggingface.co/}} implementation of \texttt{SciBERT-cased}. Our codebase for experimentation and analysis has been shared\footnote{\url{https://github.com/sayarghoshroy/ILCiteR}}.

\section{Conclusion}
\label{sec:concl}

In this work, we introduced the task of evidence-grounded local citation recommendation, with an explicit focus on recommendation interpretability. We contribute a novel dataset with over 200,000 evidence spans and sets of cited paper and support pairs, covering three subtopics within Computer Science. Our proposed recommendation system, ILCiteR, solely uses distant supervision from a dynamic evidence database and pre-trained Transformer-based language models, requiring \textit{no} explicit training.
Our findings demonstrate how a conditional neural rank ensembling approach, which utilizes semantic similarity-based ranks for lengthier queries, significantly improves downstream paper recommendation performance over purely lexical and semantic similarity based retrieval and naive rank ensembling techniques. In future, we plan to study the effects of jointly considering distant supervision from an evidence database and a secondary database of claims, findings, contributions, and named entities extracted from scientific papers for the evidence-grounded local citation recommendation task.

\section{Acknowledgements}
\label{sec:ack}

Research was supported in part by US DARPA KAIROS Program No. FA8750-19-2-1004 and INCAS Program No. HR001121C0165, National Science Foundation IIS-19-56151, and the Molecule Maker Lab Institute: An AI Research Institutes program supported by NSF under Award No. 2019897, and the Institute for Geospatial Understanding through an Integrative Discovery Environment (I-GUIDE) by NSF under Award No. 2118329.

\nocite{*}
\section{Bibliographical References}\label{sec:reference}

\bibliographystyle{lrec-coling2024-natbib}
\bibliography{lrec-coling2024-example}

\end{document}